\documentclass[
]{ceurart}

\usepackage{listings}
\usepackage{amsmath} 
\usepackage{graphicx} 
\usepackage[capitalize]{cleveref} 

\begin{document}

\copyrightyear{2026}
\copyrightclause{Copyright for this paper by its authors.
  Use permitted under Creative Commons License Attribution 4.0
  International (CC BY 4.0).}

\conference{}

\title{Designing Trustworthy LLM-based Wellbeing Recommendation through Controllable Interaction}

\author[1]{Alan Said}[orcid=0000-0002-2929-0529,email=alansaid@acm.org]
\author[1]{Alexandra Weilenmann}[orcid=0000-0002-3989-4588,email=alexandra.weilenmann@gu.se]
\address[1]{University of Gothenburg, Gothenburg, Sweden}

\begin{abstract}
Large language models (LLMs) are increasingly used to generate personalized guidance in wellbeing contexts such as physical activity, stress management, and mental health support, enabling fluent and context-aware interaction but relying on largely implicit mechanisms that shape how recommendations are expressed and adapted. We argue that this reliance on implicit adaptation through prompting and alignment limits control over guidance, responsibility framing, and user influence, which is particularly problematic in wellbeing settings where recommendations affect users’ actions and long-term outcomes. We propose a system-level perspective in which conversational behavior is structured through explicit interaction constraints, including guidance strategies, explanation styles, degrees of directness, and mechanisms for user control. Building on prior work on tangible recommendations, we show how these constraints address key challenges in wellbeing-oriented recommendation, namely trust calibration, intent alignment, and consequence awareness. We outline a modular architecture for controllable LLM-based recommendation and discuss how different configurations can be systematically designed and evaluated in relation to user-centered outcomes such as self-efficacy, perceived agency, and appropriate reliance. This paper contributes a system-level framework for designing LLM-based recommender systems that are adaptive while remaining transparent, controllable, and aligned with human wellbeing.
\end{abstract}
\begin{keywords}
Large Language Models\sep
Recommender Systems\sep
Wellbeing\sep
Interaction Design\sep
Controllability\sep
Transparency\sep
Trust\sep
Personalization
\end{keywords}

\maketitle

\begingroup
\renewcommand\thefootnote{}\footnotetext{%
  \hspace{-0.2em}\raisebox{5pt}{%
    \begin{minipage}[t]{\columnwidth}
      \footnotesize
      This is the author's version of the work. It is posted here for your personal use. Not for redistribution. The definitive Version of Record was accepted for publication in the \textit{LLM4WellRec Workshop, co-located with the 34th ACM Intl. Conference on User Modeling, Adaptation and Personalization, Gothenburg, Sweden}
    \end{minipage}%
  }%
}
\endgroup

\section{Introduction}
\label{sec:intro}
Large language models are increasingly integrated into recommender systems and user-facing applications, enabling systems that can generate fluent, context-aware, and adaptive guidance in real time. In wellbeing-oriented domains such as physical activity, stress management, and mental health support, this shifts recommendation from selecting items to producing guidance through interaction. Systems no longer present suggestions as static outputs, but engage users in ongoing dialogue, shaping how recommendations are interpreted and acted upon.

This shift brings new opportunities and changes the nature of the problem. In traditional recommender systems, outputs are typically informational or transactional. When applied to wellbeing, recommendations influence actions, routines, and experiences over time. 
As argued in prior work on tangible recommendations, such systems operate in embodied settings where consequences are physical, emotional, and social, and where users must interpret and negotiate system guidance in relation to their own state and context \cite{saidEarlyExplorationsRecommender2026}. In these settings, challenges related to trust, intent alignment, and consequence awareness become central to system design.

Current approaches to LLM-based recommendation rely heavily on implicit adaptation. Prompting, fine-tuning, and alignment techniques make it possible to generate context-sensitive responses, but provide limited control over how guidance is expressed, how strongly it is framed, or how responsibility is communicated. As a result, interactional properties emerge from the generative process, not from explicit specification. This makes them difficult to inspect, compare, or systematically vary across system designs. What is gained in flexibility is often lost in controllability \cite{amershiGuidelinesHumanAIInteraction2019,weidingerTaxonomyRisksPosed2022}.

This lack of control is problematic in wellbeing settings. The same recommendation may be motivating in one situation and inappropriate in another, even for the same user. A system that suggests increased activity may motivate one user while pushing another beyond their limits. These effects are not captured by standard notions of relevance or accuracy, but emerge through interaction.

Research in human–AI interaction has long emphasized the importance of transparency, controllability, and user agency for appropriate reliance \cite{leeTrustAutomationDesigning2004,shneidermanHumanCenteredArtificialIntelligence2020,schemmerAppropriateRelianceAI2023}. However, these properties are difficult to realize in LLM-based systems, where behavior is dynamically generated and context-dependent. Existing work on explainable AI, including recent work on LLM-based explanations, suggests that post hoc explanations alone are insufficient when the underlying behavior remains opaque \cite{saidExplainingRecommendationsLarge2025}. Making systems understandable requires explaining outputs and structuring how they are produced.

In this paper, we argue that LLM-based recommendation for wellbeing should be approached as an AI systems design problem. We propose to model conversational behavior through explicit and configurable interaction constraints, not as an emergent property of generation. These constraints include, for example, how guidance is phrased, how direct it is, how explanations are provided, and how users can influence or override system behavior.

Interaction constraints provide a systematic way to address these challenges. We outline a modular architecture in which interactional properties are implemented as system-level components layered on top of large language models, supporting a shift from implicit adaptation to inspectable and comparable system designs.

The contributions of this paper are threefold. First, we identify limitations of current LLM-based recommendation approaches in wellbeing contexts, focusing on the role of implicit adaptation and limited controllability. Second, we introduce a framework for structuring conversational behavior through explicit interaction constraints, grounded in the dimensions of trust, intent, and consequence. Third, we outline an architectural and evaluation perspective that links system-level design choices to user-centered outcomes such as self-efficacy, perceived agency, and appropriate reliance. Together, these contributions provide a foundation for designing LLM-based recommender systems that are adaptive, but also transparent, controllable, and aligned with human wellbeing.

\section{Related Work}
Research on recommender systems for health and wellbeing has grown substantially in recent years \cite{elsweilerEngenderingHealthRecommender2016,hauptmannResearchDirectionsRecommender2022}, reflecting a broader shift toward applications that aim to influence behavior, support self-management, and improve long-term outcomes. Early work in health recommender systems (HRS) emphasizes their potential to motivate behavior change and provide actionable guidance \cite{schaferHealthAwareRecommender2017}, but also highlights significant variation in how such systems are designed and evaluated \cite{croonHealthRecommenderSystems2021}. As shown in a systematic review of HRSs, many systems focus on delivering lifestyle, nutrition, or health information recommendations, yet often rely on limited evaluation protocols or narrow notions of effectiveness. More recent work argues that health recommender systems introduce distinct challenges, including the need to account for trust, privacy, and user involvement in decision-making \cite{roySystematicReviewResearch2022}.

These challenges are closely related to a broader shift in recommender systems research toward applications for social good. 
Instead of optimizing for engagement or accuracy alone, recommender systems are increasingly expected to contribute to societal goals such as wellbeing, fairness, and sustainability \cite{jannachRecommenderSystemsGood2025,saidRecommenderSystemsSocial2024}. This shift requires rethinking both system objectives and evaluation practices, including the need for longitudinal and human-centered evaluation approaches that capture real-world impact. At the same time, work on fairness in healthcare recommendation highlights that core concepts such as fairness and equity are context-dependent and difficult for users to interpret, reinforcing the need for systems that are transparent and sensitive to situational factors \cite{keckiUnderstandingFairnessRecommender2024}.

A parallel line of work in human-centered AI and interaction design emphasizes that user experience, meaning, and agency emerge through situated interaction, not solely from system functionality. Studies of technology use in everyday and wellbeing contexts show that how systems are interpreted and integrated into practice depends on interactional context, local meaning-making, and user control \cite{weilenmannSelfiesWildStudying2020,cernaSupportingSelfmanagementRadiationinduced2019}. This perspective, developed in interaction design and HCI research, highlights that properties such as agency, ownership, and responsibility must be designed into systems, not assumed to arise from adaptive behavior.
In the context of personalization and recommender systems, this aligns with prior work arguing that recommendations affecting physical activity and wellbeing must be understood as part of embodied and situated practices \cite{saidEarlyExplorationsRecommender2026}.

Large language models introduce new possibilities within this landscape by enabling conversational, adaptive, and context-aware recommendation. Recent work has explored their use for explanation generation \cite{silvaLeveragingChatGPTAutomated2024}, conversational interfaces \cite{sunTrustInterfaceHow2024}, and richer user modeling \cite{chenWhenLargeLanguage2024}. However, existing research also points to limitations of LLM-based approaches. In particular, the fluency and plausibility of generated explanations do not necessarily improve user understanding or support appropriate trust \cite{saidExplainingRecommendationsLarge2025}. More broadly, language models are associated with risks related to opacity, unpredictability, and limited controllability of outputs \cite{weidingerTaxonomyRisksPosed2022}, raising concerns about their use in high-stakes domains such as healthcare and wellbeing.

Efforts to address these challenges have often focused on explainable AI and transparency mechanisms. In healthcare, explainability is seen as essential for building trust and enabling users and professionals to understand and apply system outputs \cite{yangExplainableArtificialIntelligence2022}. However, explanation is typically treated as an additional layer on top of underlying system behavior. As a result, even well-designed explanations may fail to provide meaningful control or understanding when the behavior itself remains implicit or difficult to inspect.

Put together, existing work highlights three key gaps. First, wellbeing-oriented recommender systems require attention to trust, agency, and long-term consequences, but these aspects are not systematically integrated into system design. Second, LLM-based approaches increase flexibility and expressiveness, but do not resolve issues of control and transparency. Third, current approaches to explainability focus on outputs rather than on structuring the interactional properties that shape how recommendations are experienced.

In contrast to prior work, this paper adopts a system-level perspective in which conversational behavior is treated as an explicit design space. 
We propose to model interaction through configurable constraints that govern how recommendations are expressed and how users can engage with them. This approach builds on insights from recommender systems, human–AI interaction, and wellbeing research, while addressing a gap in current work on how LLM-based recommendation can be made controllable, inspectable, and aligned with human wellbeing

\section{Wellbeing Recommendation as Interaction Design}
\label{sec:id}
Recommender systems in wellbeing settings are often framed as extensions of traditional personalization, where improved user modeling leads to better recommendations. This assumes that the primary challenge is identifying what should be recommended given a user’s preferences, state, or goals. However, this framing becomes limited when recommendations influence behavior instead of consumption. In wellbeing-oriented applications, recommendations are not simply selected and consumed. They are interpreted, negotiated, and sometimes resisted. Prior work shows that guidance related to physical activity, lifestyle, or health is experienced in relation to users’ own knowledge, expectations, and bodily awareness \cite{saidEarlyExplorationsRecommender2026}. The same recommendation may be motivating in one situation and inappropriate in another, even for the same user. What matters is both what is suggested and how it is understood in context. The focus therefore shifts from recommendation as output to recommendation as interaction.

A similar view is reflected in work in interaction design and human–computer interaction, which emphasizes that meaning and agency emerge through situated use, not solely from system functionality. Technologies are incorporated into everyday practices, where users actively interpret and adapt system behavior in relation to their own goals and constraints \cite{weilenmannSelfiesWildStudying2020, cernaSupportingSelfmanagementRadiationinduced2019}. In these contexts, systems are embedded in routines, habits, and competing priorities, and their effects unfold over time, not within isolated interactions. As a result, properties such as trust, control, and responsibility are not fixed system characteristics, but are shaped through ongoing interaction. Building on this perspective, prior work identifies three key dimensions of recommendation in embodied settings: trust and interpretation, intent and alignment, and consequence awareness \cite{saidEarlyExplorationsRecommender2026}. These dimensions point to the limits of applying conventional recommender system logic in domains where recommendations shape action, not just choice.

Trust and interpretation concern how users make sense of recommendations and decide whether to act on them. Trust does not follow automatically from accuracy, but is shaped by how guidance is framed, how confident it appears, and how well it aligns with users’ expectations. In conversational systems, this becomes more pronounced, as fluent and confident responses may be interpreted as authoritative even when they are uncertain or inappropriate. Intent and alignment address the difficulty of representing user goals in contexts where they are fluid and situation-dependent. Users may shift between competing goals such as exertion and recovery, or between short-term constraints and long-term aspirations, and systems that rely on static or inferred representations risk misalignment. Consequence awareness extends the scope further by emphasizing that recommendations in wellbeing domains have effects that accumulate over time, including physical outcomes, emotional responses, and changes in routine. These effects are rarely captured by standard evaluation metrics, and systems typically lack mechanisms for representing or responding to them.

Collectively, these dimensions point to a different problem formulation. The challenge is to identify relevant recommendations and to shape how they are expressed and integrated into interaction. This becomes more pronounced in LLM-based systems, where recommendations are generated dynamically through conversation. The flexibility of these systems allows for adaptation in tone, framing, and explanation, but also introduces variability and opacity, as small changes in phrasing can significantly affect how guidance is perceived. 
Treating recommendation as an interaction design problem instead of a prediction problem therefore requires mechanisms to structure these interactional properties explicitly. In the next section, we introduce interaction constraints as a way to make these properties configurable, inspectable, and open to systematic design and evaluation.

\section{Interaction Constraints for Controllable LLM Behavior}
The previous section argued that wellbeing recommendation should be understood as an interaction design problem, where how guidance is expressed is as important as what is recommended. This raises a practical question: how can such interactional properties be systematically designed, instead of being left to emerge implicitly from model behavior?

Current LLM-based systems rely primarily on prompting, fine-tuning, and alignment techniques to shape responses. While these approaches can influence outputs, they provide limited control over specific interactional properties and offer few guarantees regarding consistency or interpretability. Small changes in prompts or context may lead to substantial variation in how guidance is phrased, how confident it appears, or how responsibility is framed. As a result, key aspects of interaction remain implicit, making them difficult to specify, compare, or evaluate across system designs.

To address this limitation, we propose treating conversational behavior as structured through interaction constraints. These constraints are explicit, configurable parameters that govern how recommendations are expressed and how users can engage with them. Constraints make these properties visible at the system level, supporting controlled variation and systematic design, without relying solely on the generative model to determine them.

We focus on four types of constraints that are relevant to wellbeing. As shown in \cref{tab:interaction-constraints}, these constraints shape how recommendations are experienced, particularly in terms of trust, agency, and interpretation. For example, a physical activity recommendation can be expressed as ``You should go for a run today,'' reflecting high directness and system-centered responsibility framing, or as ``You might consider a light run today if you feel up to it,'' reflecting lower directness and a more user-centered framing. These alternatives represent different  constraint configurations and are likely to lead to different interpretations, levels of trust, and user responses.

\begin{table}[t]
\centering
\caption{Interaction constraints in LLM-based wellbeing recommendation}
\label{tab:interaction-constraints}
\begin{tabular}{p{2.6cm} p{4.2cm} p{3.4cm} p{3.9cm}}
\toprule
\textbf{Constraint} & \textbf{Description} & \textbf{Example realizations} & \textbf{Implications} \\
\midrule
Guidance strategy & How recommendations are positioned relative to the user & Suggestion vs.\ directive phrasing & Shapes perceived authority and autonomy \\

Degree of directness & Strength and explicitness of the recommendation & Tentative (``might'') vs.\ assertive (``should'') & Influences compliance and trust calibration \\

Explanation style & How reasoning is communicated & Data-driven, general, or no explanation & Affects understanding and perceived transparency \\

Responsibility framing \& user control & How decision-making authority is distributed and adjusted & User overrides, goal adjustment, alternative suggestions & Supports agency but may increase user burden \\
\bottomrule
\end{tabular}
\end{table}

First, guidance strategy defines how recommendations are positioned in relation to the user. A system may frame its output as a suggestion, a recommendation, or a directive. These alternatives differ in tone and in how they allocate responsibility and influence user behavior. In wellbeing settings, presenting guidance as a suggestion, not an instruction, may support autonomy and reduce the risk of over-reliance.

Second, degree of directness captures how explicitly the system communicates its recommendation. Responses may range from tentative (``you might consider...'') to assertive (``you should...''). Directness affects both interpretive trust and user action. Highly direct recommendations may be persuasive but risk being inappropriate in uncertain contexts, whereas more tentative phrasing may better reflect uncertainty but reduce perceived usefulness.

Third, explanation style concerns how and whether the system provides reasons for its recommendations. Explanations may be based on user data, general knowledge, or contextual cues, and may vary in level of detail and specificity. Prior work shows that explanation influences trust and decision-making, but explanations that are fluent without being grounded may create a false sense of understanding \cite{saidExplainingRecommendationsLarge2025}. Making explanation style an explicit constraint allows systems to vary and evaluate how different forms of explanation affect user interpretation.

Fourth, responsibility framing and user control define how decision-making authority is distributed between system and user. This includes both how the system communicates responsibility (``this is based on your recent activity'' vs. ``this is a general recommendation'') and what mechanisms users have to influence system behavior. Examples include the ability to adjust goals, override recommendations, or request alternative suggestions. Such mechanisms are critical for supporting agency and appropriate reliance, particularly in contexts where recommendations affect personal wellbeing.

These constraints are not independent. They interact in shaping how recommendations are perceived in practice. For example, a highly direct recommendation combined with strong system-centered responsibility framing may encourage compliance, while a tentative recommendation with user-centered framing may invite reflection. Treating these properties as explicit parameters makes it possible to explore such combinations systematically, rather than leaving them to emerge unpredictably from generation.

Importantly, these constraints do not replace adaptation, but structure it. The goal is to provide a layer in which key interactional properties can be specified, inspected, and varied, not to restrict the flexibility of LLMs. This shifts the focus from implicit, prompt-driven behavior to configurable system design, where different interaction strategies can be implemented and compared.

By making interactional properties explicit, these constraints provide a foundation for linking system behavior to user-centered outcomes. They enable questions that are difficult to address in current systems: how does the degree of directness affect trust calibration? How does responsibility framing influence perceived agency? How do different explanation styles shape understanding and reliance? In the next section, we build on this by outlining a system and evaluation perspective in which such questions can be addressed in a principled and reproducible manner.

\section{System and Evaluation Perspective}
The notion of interaction constraints provides a way to make conversational behavior explicit. To operationalize this idea, it needs to be grounded in both system design and evaluation. This section outlines a perspective in which these constraints are treated as system-level components, and where their effects can be systematically studied in relation to user-centered outcomes.

At the system level, the key idea is to separate generation from interaction control. Current LLM-based systems typically rely on a single generative process, shaped indirectly through prompts and context. In contrast, we consider an architecture in which the language model is embedded within a modular structure that includes an explicit interaction layer. This layer governs how responses are constructed, using interaction constraints to regulate properties such as guidance strategy, directness, explanation, and responsibility framing.

Conceptually, the architecture consists of three components. A generation component, based on a large language model, produces candidate responses given user input and contextual information. An interaction policy layer applies constraints to shape these responses, selecting or modifying outputs according to the desired interaction configuration. Finally, a user modeling and context component provides information about user state, preferences, and situational factors, which can inform both generation and constraint selection. This separation means conversational behavior can be configured independently of the underlying model, supporting controlled variation across system designs.

This structure supports a different approach to evaluation. 
Moving away from treating the system as a monolithic black box, individual interactional properties can be varied and compared. For example, systems can be instantiated with different levels of directness or alternative responsibility framings, while keeping other factors constant. This makes it possible to isolate the effects of specific design choices on user experience and behavior.

Such an approach is particularly important in wellbeing settings, where standard evaluation metrics are insufficient. Traditional recommender system metrics, such as accuracy or engagement, capture only a narrow view of system performance. In contrast, prior work emphasizes the need to consider outcomes such as self-efficacy, perceived agency, trust, and appropriate reliance \cite{leeTrustAutomationDesigning2004, saidEarlyExplorationsRecommender2026}. These constructs capture how users understand and act on recommendations, instead of focusing on whether they simply interact with them.

Evaluating these outcomes requires a combination of methods. Controlled experiments can be used to compare different system configurations, measuring how variations in interaction constraints influence user responses. Validated psychometric instruments can assess constructs such as self-efficacy and trust, while interaction-level data can provide insight into how users engage with and adapt to system behavior over time. Qualitative methods, including interviews and interaction analysis, can further capture how recommendations are interpreted in context.

Importantly, this perspective shifts evaluation from optimizing a single objective to understanding trade-offs. For instance, more direct recommendations may increase short-term compliance but reduce perceived agency, while more tentative guidance may support reflection but decrease action. Similarly, explanation styles that increase perceived transparency may also lead to over-reliance if not carefully calibrated. By framing these as design variables instead of side effects, systems can be evaluated based on how they balance competing outcomes.

This also encourages more longitudinal evaluation. Over time, the effects of recommendations accumulate, and short-term interaction metrics may not reflect long-term impact. A system that encourages sustained, appropriate engagement may be preferable to one that maximizes immediate response. By linking interaction constraints to user-centered outcomes, the proposed approach provides a foundation for studying such effects in a structured way.

Overall, this system and evaluation perspective enables a shift from implicit, model-driven behavior to explicit, inspectable system design. It allows researchers and practitioners to move beyond treating LLM-based recommenders as black boxes, and instead to study how specific interactional choices influence user experience and outcomes. This is a necessary step toward designing systems that are both adaptive and also aligned with the goals of trustworthiness and wellbeing.

\section{Discussion and Conclusion}
This work argues for a shift in how LLM-based recommender systems are conceptualized in wellbeing contexts. Our approach frames recommendation as an interaction design problem rather than a prediction task, with the expression of guidance, the allocation of responsibility, and the structure of user engagement shaping system behavior. This shift is not merely conceptual. It reflects the realities of systems that influence users’ actions, routines, and experiences over time, where outcomes depend both on how recommendations are delivered and on the content of those recommendations.

The introduction of interaction constraints provides a way to operationalize this perspective. By making properties such as guidance strategy, directness, explanation, and responsibility framing explicit, systems can move from implicit, model-driven behavior toward configurable and inspectable interaction design. 
This supports a more systematic exploration of how conversational behavior affects trust, agency, and decision-making. It also aligns with broader calls in human-centered AI to design for appropriate reliance rather than maximizing compliance or engagement \cite{leeTrustAutomationDesigning2004}.

At the same time, this view highlights limitations in current approaches to LLM-based recommendation. The flexibility and fluency of language models make them well-suited for adaptive interaction, but also introduce risks. Responses may appear authoritative without being grounded, adaptivity may obscure underlying assumptions, and users may over-rely on system guidance in contexts where judgment should remain with the user. These risks are amplified in wellbeing domains, where recommendations can have cumulative physical, emotional, and behavioral consequences. Addressing them requires  better models as well as more explicit control over how those models are used in interaction.

More broadly, the proposed approach points toward a reconfiguration of evaluation practices. If system behavior is understood in terms of interaction, then evaluation must move beyond accuracy and engagement to consider user-centered outcomes like self-efficacy, perceived agency, and trust. These outcomes are not easily reduced to single metrics, and often involve trade-offs that must be made explicit. Designing for wellbeing thus involves balancing competing objectives instead of optimizing a single target.

Several open questions remain. One challenge lies in how interaction constraints should be specified and adapted in practice, particularly in dynamic contexts where user needs and preferences change over time. Another concerns how to integrate user control in meaningful ways, without placing undue burden on users or undermining system support. Finally, there is a need for longitudinal studies that examine how different interaction strategies affect behavior and wellbeing over extended periods, beyond isolated interactions.

In conclusion, this paper contributes a system-level perspective on LLM-based recommendation for wellbeing, centered on the design of interaction rather than the optimization of outputs. By introducing interaction constraints as a means of structuring conversational behavior, we provide a foundation for developing systems that are adaptive, transparent, controllable, and aligned with human agency. This opens up new directions for research at the intersection of recommender systems, human–AI interaction, and wellbeing, where the goal is not simply to recommend effectively, but to support users in acting, reflecting, and maintaining ownership over their decisions.

\section*{Declaration on Generative AI}
During the preparation of this work, the author(s) used ChatGPT and Grammarly for Grammar and
spelling check and rephrasing. After using these tool(s)/service(s), the author(s) reviewed and edited the content as needed and take(s) full responsibility for the publication’s content.

\bibliography{sample}

\end{document}